\documentclass[prd,preprint,showpacs,amssymb,superscriptaddress,nofootinbib]{revtex4}

\usepackage{amsmath}
\usepackage{amsfonts}
\usepackage{amssymb}
\usepackage{graphicx}
\usepackage{color}
\usepackage{graphicx}
\usepackage{epstopdf}%

\renewcommand{\Im}{\mathrm{Im}}
\renewcommand{\Re}{\mathrm{Re}}

\def\slash#1{\setbox0=\hbox{$#1$}  
  \dimen0=\wd0     
  \setbox1=\hbox{/} \dimen1=\wd1  
  \ifdim\dimen0>\dimen1   
     \rlap{\hbox to \dimen0{\hfil/\hfil}} 
     #1     
  \else     
     \rlap{\hbox to \dimen1{\hfil$#1$\hfil}} 
     /      
  \fi}      %

\newcommand{\Dn}{\slash{n} }
\newcommand{\Dbn}{\slash{\bar n}}

\begin{document}

\setcounter{secnumdepth}{5}

\vglue 5mm

\title{Study of Two-Photon Corrections in the $p\bar{p} \rightarrow e^+e^-$ Process: Hard Rescattering Mechanism}
%
\author{Julia Guttmann}
\affiliation{Institut f\"ur Kernphysik, Johannes-Gutenberg Universit\"at, D-55099 Mainz, Germany}

\author{Nikolai Kivel}
\affiliation{Institut f\"ur Kernphysik, Johannes-Gutenberg Universit\"at, D-55099 Mainz, Germany}
\affiliation{Helmholtz Institut Mainz, Johannes-Gutenberg Universit\"at, D-55099 Mainz, Germany}
\affiliation{Petersburg Nuclear Physics Institute, Gatchina, 188350, Russia}

\author{Marc Vanderhaeghen}
\affiliation{Institut f\"ur Kernphysik, Johannes-Gutenberg Universit\"at, D-55099 Mainz, Germany}

\date{\today}
\begin{abstract}
We investigate the two-photon corrections to the process $p\bar{p} \to e^+e^-$ at large momentum transfer, aimed to access the time-like nucleon form factors. We estimate the two-photon corrections using a hard rescattering mechanism, which has already been used to calculate the 
corresponding corrections to elastic electron-proton scattering.
Using different nucleon distribution amplitudes, we find that the two-photon corrections 
to the $p\bar{p} \to e^+e^-$ cross sections in the momentum transfer range 5 - 30 GeV$^2$ 
is below the 1 \% level. 

\end{abstract}
%
\maketitle 

\section{Introduction}

Electromagnetic form factors (FFs) provide important information on the structure of the nucleon. 
Consequently, there has been much effort in their measurement both in the space-like as well as in the time-like regions. 

The space-like electromagnetic FFs, which provide information on spatial distributions of quarks in the nucleon,  can be investigated in elastic electron-proton scattering; 
for recent reviews see {\it e.g.}  
Refs.~\cite{HydeWright:2004gh,Arrington:2006zm,Perdrisat:2006hj}. 
Two experimental methods exist to extract the ratio of electric ($G_E$) to magnetic ($G_M$) 
proton FFs. 
Historically, the first method involves unpolarized measurements employing the Rosenbluth separation technique, which gives a direct acess to the space-like FFs through the slope and intercept of the $\varepsilon$-dependence of the cross section in the one-photon ($1 \gamma$) exchange approximation~:
\begin{equation}
d\sigma =\mathcal{C}(Q^2,\varepsilon)\Big[G_M^2(Q^2) + \frac{\varepsilon}{\tau}G_E^2(Q^2)\Big],
\end{equation}
where $\varepsilon$ and $Q^2$ are the virtual photon polarization parameter and virtuality respectively, 
$\tau = Q^2 / 4 m_N^2$, with $m_N$ the nucleon mass, and where $\mathcal{C}$ is a known 
phase space factor.  
In more recent years, 
polarization experiments using polarized electron beams 
on polarized targets or measuring the recoil nucleon polarization, in the elastic $e-p$ scattering, 
provided another experimental way to access $G_E/G_M$. 
The ratio of polarization of the recoiling proton perpendicular to its motion ($P_t$) to polarization along its motion ($P_l$) is directly related to the ratio of electric to magnetic proton FFs 
\begin{equation}
\frac{P_t}{P_l} = - \sqrt{\frac{2\varepsilon}{\tau(1+\varepsilon)}}\,\frac{G_E(Q^2)}{G_M(Q^2)}.
\end{equation}
Such polarization experiments have been performed for momentum transfers up to 
$8.5$~GeV$^2$ to date and have shown that the ratio of the electric to magnetic proton FFs is decreasing with increasing momentum transfer \cite{Jones:1999, Punjabi:2005wq, Gayou02, Puckett:2010ac}. This finding is in contrast to the well known scaling-behavior of $G_E/G_M$ determined by the Rosenbluth separation technique. The discrepancy between data of unpolarized Rosenbluth measurements and polarization experiments has triggered a whole new field studying the influence of two-photon ($2 \gamma$) exchange corrections~\cite{Guichon:2003, Blunden:2003, YCC04, Arrington:2007ux, Borisyuk:2008db, Kivel:2009eg}; see Ref.~\cite{Carlson:2007sp} for a recent review and references therein. 
The finding of those works is that two-photon exchange corrections to the Rosenbluth cross section are a possible explanation for the discrepancy, whereas the $2 \gamma$ exchange effects do not impact the polarization transfer extraction of $G_E/G_M$ in a significant way.
Recently, a first empirical extraction of the three $2 \gamma$-exchange amplitudes to elastic electron-proton scattering has been performed~\cite{Guttmann:2010au} 
based on measurements of cross sections~\cite{Qattan:2004ht} 
and polarization observables~\cite{Meziane:2010} at a common value of four-momentum transfer, around $Q^2 = 2.5$~GeV$^2$. It confirms that a common description of unpolarized measurements and of polarization observables invokes empirical $2 \gamma$-amplitudes with relative magnitude up to about 3~\%. 

The measurements of nucleon FFs at space-like momentum transfers, through
elastic electron-nucleon scattering, are complemented by measurements in the
time-like region, through the crossed processes $p\bar{p} \to e^+e^-$ and $e^+e^-\to N \bar N$, which
access the vector mesonic excitation spectrum of hadrons. 
The latter process has been measured in recent years at $e^+ e^-$ facilities,
such as Da$\Phi$ne, CLEO,  and BABAR.  These measurements have revealed that the
nucleon FFs at time-like momentum transfers are
significantly larger than their space-like counterparts when considering the same magnitude 
for the virtuality $q^2 = - Q^2$. In particular, for momentum
transfers with magnitude around 10 GeV$^2$, the time-like FFs were found to be enhanced 
by a factor of two. New measurements are planned in
the near future at BES-III and at PANDA@FAIR, bringing time-like (positive) $q^2$ values around 20~GeV$^2$ into reach. 
Such new measurements will explore the at present still largely uncharted time-like region in much greater detail and complement our picture of the nucleon.

The time-like FFs are complex quantities due to the interactions of the hadrons in the initial and final state, respectively. Their absolute values can be determined from measurements of the angular distribution of the unpolarized c.m. cross section in the $1 \gamma$-approximation~:
\begin{equation}
d\sigma_{\mathrm{c.m.}, 1\gamma} = \mathcal{C}(q^2)\,\left[|G_M|^2(1 + \cos^2\theta) + \frac{1}{\tau} |G_E|^2\sin^2\theta \right],
\end{equation}
whereas the phases are related to polarization observables. Since $2 \gamma$-exchange plays a crucial role in the extraction of electromagnetic FFs in the space-like region, investigating its influence in the time-like region seems to be an obvious task. Even though some theoretical works have been done \cite{Gakh:2005,Gakh:2005b}, there are no comparable calculations so far to estimate the 
$2 \gamma$-exchange corrections for the corresponding time-like processes.

In this work we investigate the $2 \gamma$-exchange corrections to the 
process $p\bar{p} \to e^+e^-$ at large momentum transfer $q^2$. To provide a first estimate of the corrections we consider a perturbative QCD (pQCD) factorization approach, 
which has already been used to calculate the 
corresponding corrections to elastic electron-proton scattering~\cite{Kivel:2009eg}.

The paper is organized as follows: The general formalism including $2 \gamma$-exchange is presented in Section \ref{sec:observ}. In Section \ref{sec:calc} we estimate the hard $2 \gamma$-exchange contribution at large momentum transfers by relating the $2 \gamma$-exchange amplitude to the leading twist nucleon distributions. The results of the calculation are discussed in Section \ref{sec:results}. Some concluding remarks are given in Section \ref{sec:conclusion}.

\section{General expression of the Observables inlcuding 2$\mathbf{\gamma}$-exchange}
\label{sec:observ}

In order to describe the annihilation of a proton and an antiproton into a lepton pair,
\begin{equation}
p(p_1, \lambda_{N_1}) + \bar{p}(p_2, \lambda_{N_2}) \to l^-(k_1, h_1) + l^+(k_2,h_2),
\end{equation}
where $\lambda_{N_1}$, $\lambda_{N_2}$, $h_1$ and $h_2$ are helicities of the nucleons and leptons respectively, we adopt the definitions
\begin{eqnarray}
P=\frac{p_1 - p_2}{2},  \qquad K=\frac{k_1 - k_2}{2},  \qquad q^2=(p_1+ p_2)^2, 
\end{eqnarray}
and the Mandelstam variables
\begin{equation}
s = q^2= (p_1 + p_2)^2, \quad t= (p_1 - k_2)^2, \quad u=(p_1 -k_1)^2.
\end{equation}
The process can be described by two independent kinematical invariants, which we choose as the variables $q^2$ and $t$.

The amplitude of the reaction is related by crossing to the corresponding scattering amplitude for elastic electron-proton scattering. Neglecting the lepton masses, the matrix element including multi-photon exchange is parameterized by three generalized form factors. Several equivalent representations exist. Here we use the representation, which was first introduced in Ref.~\cite{Guichon:2003}. The matrix element can be written in the form
\begin{equation}
 T = \frac{e^2}{q^2}\Bigg\{\bar{u}(k_2,h)\gamma_\mu v(k_1,- h)\times\bar{v}(p_2, \lambda_{N_2})\left[\tilde G_M\gamma^\mu - \tilde F_2\,\frac{1}{m_{N}}P^\mu + \tilde F_3\,\frac{1}{m_{N}^2}P^\mu\slash{K} \right]u(p_1, \lambda_{N_1})\Bigg\},
\label{eq:tmatrix}
\end{equation}
where  $\tilde G_M$, $\tilde F_2$ and $\tilde{F}_{3}$ are complex functions of $q^2$ and $t$. Neglecting the lepton masses implicates that the outgoing electron and positron have opposite helicities.

In the following, we also use the generalized form factor
\begin{equation}
\tilde G_E = \tilde G_M - \Big(1-\frac{q^2}{4m_{N}^2}\Big) \tilde F_2. 
\end{equation}
In the Born approximation $\tilde G_M$ and $\tilde G_E$ reduce to the usual proton FFs and do not depend on $t$, while $\tilde F_3$ vanishes. In order to identify the $1 \gamma$ and 
$2 \gamma$-exchange contributions, we introduce the decompositions
\begin{eqnarray}
\tilde G_M(q^2,t) &\equiv& G_M(q^2) + \delta\tilde G_M(q^2, t), \nonumber \\
\tilde G_E(q^2,t) &\equiv& G_E(q^2) + \delta\tilde G_E(q^2, t), \nonumber \\
\tilde F_3(q^2,t) &\equiv& \delta\tilde F_3(q^2, t).
\end{eqnarray}
$G_M$ and $G_E$ are the time-like proton magnetic and electric FFs and $\tilde F_3$, $\delta\tilde G_M$ and $\delta\tilde G_E$ are amplitudes of order $e^2$, which originate from processes involving the exchange of at least two photons.

To compute the differential cross section of the reaction, we use the center-of-mass (c.m.) frame, where the momenta of the incoming proton and antiproton have opposite directions. In this frame the variable $t$ can be related to the c.m.-scattering angle $\theta$ between the incident proton and the outgoing electron.
Calculating the cross section up to next order of $e^2$ leads to the expression
\begin{eqnarray}
 d\sigma_{\mathrm{c.m.}} &=& \mathcal{C}\big(q^2\big)\Bigg[|G_M|^2(1 + \cos^2\theta) + \frac{1}{\tau}|G_E|^2\sin^2\theta \nonumber \\
& & +\, 2\Re[G_M\delta\tilde{G_M}^*](1 + \cos^2\theta) +\, 2\frac{1}{\tau}\Re[G_E\delta\tilde{G_E}^*]\sin^2\theta \nonumber \\
& & +\, 2\Big(\Re[G_M \tilde{F_3}^*]- \frac{1}{\tau}\Re[G_E \tilde{F_3}^*]\Big)\sqrt{\tau(\tau -1)}\cos\theta\sin^2\theta \Bigg],
\label{eq:crosssec}
\end{eqnarray}
with
\begin{equation}
\tau = \frac{q^2}{4m_{N}^2}, \quad \mathcal{C}(q^2) = \frac{e^4}{64\pi^2q^2}\sqrt{\frac{\tau-1}{\tau}}.
\end{equation}
In the $1 \gamma$-exchange approximation, only the first two terms of Eq. (\ref{eq:crosssec}) contribute to the cross section and it reduces to the well known formula of the unpolarized cross section:
\begin{equation}
d\sigma_{\mathrm{c.m.}, 1\gamma} = \mathcal{C}(q^2)\,\left[|G_M|^2(1 + \cos^2\theta) + \frac{1}{\tau} |G_E|^2\sin^2\theta \right].
 \label{cs:1photon}
\end{equation}
The other part of Eq. (\ref{eq:crosssec}) represents the interference of 
$1 \gamma$ and $2 \gamma$-exchange processes.

In order to determine the imaginary part of the time-like form factors it is necessary to study polarization observables. An observable which gives acess to the imaginary part of the electric and magnetic form factor is the single spin asymmetry when either the proton or antiproton is polarized normal to the scattering plane, which does not require polarization of the leptons in the final state. Polarization of the proton or antiproton along or perpendicular to its motion, but in the scattering plane, in contrast 
also requires a polarized lepton.

The single spin asymmetry can be defined as
\begin{equation}
A_y = \frac{d\sigma^{\uparrow} - d\sigma^{\downarrow}}{d\sigma^{\uparrow} + d\sigma^{\downarrow}},
\end{equation}
where $d\sigma^{\uparrow}$ ($d\sigma^{\downarrow}$) denotes the cross section for an incoming nucleon with positiv (negativ) perpendicular polarization.
In the case of a polarized proton the single spin asymmetry up to next order in 
$e^2$ can be obtained as
\begin{eqnarray}
A_y &=& \frac{1}{d\sigma_{\mathrm{c.m.}}}\,\frac{\mathcal{C}}{\sqrt{\tau}}\ 2\sin\theta\Bigg\{\, \Big(\Im[ G_E G_M^*] + \Im[G_E \delta\tilde G_M^*]+ \Im[\delta \tilde G_E  G_M^*]\Big)\cos\theta \nonumber \\[1.0ex]
& & +\  \sqrt{\tau(\tau-1)}\Big(\Im[G_M \tilde F_3^*]\cos^2\theta +\ \Im[G_E\tilde F_3^*]\sin^2\theta\Big)\Bigg\}.
\label{eq:ssa}
\end{eqnarray}
In contrast to space-like processes the single spin asymmetry $A_y$ in the time-like region does not vanish in the Born approximation.

\section{Calculation of the $2 \gamma$-Exchange Contribution at large $q^2$}
\label{sec:calc}

To calculate the $2 \gamma$-exchange corrections in $p \overline{p} \rightarrow e^+ e^-$ at large momentum transfers we consider a factorization approach using the concept of hadron distribution amplitudes (DAs). We follow the experience gained by the space-like process $e p \to e p$, 
for which the amplitudes $\delta \tilde G_{M}$ and $\tilde F_{3}$ were 
computed~\cite{Borisyuk:2008db, Kivel:2009eg} at large momentum transfer  $Q$ in the form of a convolution of a hard kernel $H$, which can be calculated in QCD perturbation theory, and the nonperturbative contributions $\Psi$, which can be related to the DAs of proton and antiproton, for instance:
\begin{equation}
\delta \tilde G_{M}(Q^{2},\varepsilon )\simeq \Psi*H(Q^{2},\varepsilon )*\Psi, 
\label{HS}
\end{equation}
where the asterisk denotes the convolutions with respect to the participating quark momentum fractions. This result represents the leading order contribution with respect to an expansion in $1/Q$.  The important feature of such an approach is that the virtualities of both photons must be large: $-q^{2}_{1}\sim -q^{2}_{2}\sim Q^{2}$.  The corresponding hard subprocess involves only one hard gluon exchange and is therefore suppressed by the strong coupling $\sim \alpha_{s}$. As all spectator quarks are involved in the hard scattering process described by Eq.~(\ref{HS}), we shall refer to it  as the hard rescattering contribution.  

\begin{figure}[ht]
\includegraphics[scale=0.6]{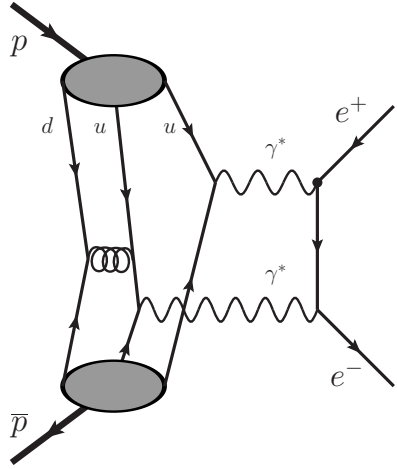}
\caption{Diagram for $p \bar{p} \rightarrow e^+ e^-$ including the exchange of two hard photons}
\label{Fig:diagrams}
\end{figure}

This mechanism can be simply generalized to the crossing channel $p \overline{p} \rightarrow e^+ e^-$ at large momentum transfer $q^{2}>0$. 
A typical diagram of the leading pQCD contribution to the $2\gamma$-exchange correction to the annihilation amplitude is illustrated in 
Fig.\ref{Fig:diagrams}.
The simple analysis allows to conclude that a description of the corresponding  hard time-like subprocess can be obtained directly from the space-like one using the crossing symmetry.  Therefore we obtain that the leading order asymptotic behavior of the time-like $2 \gamma$-exchange amplitudes can be represented by the same form as Eq.~(\ref{HS}) for the space-like process.  

In the case of $q^2,t \gg m_{N}^2$  the momenta of proton and antiproton in the c.m.-frame can be expressed by two light-like vectors $n$, $\bar{n}$:
\begin{eqnarray}
p_1 &\simeq& \sqrt{s}\,\frac{\bar{n}}{2}, \qquad \bar{n} = (1,0,0,1), \nonumber \\
p_2 &\simeq& \sqrt{s}\,\frac{n}{2}, \qquad n = (1,0,0,-1).
\label{p12}
\end{eqnarray}
The lepton momenta are defined as
\begin{eqnarray}
k_1^\mu = \bar{\eta}\frac{\sqrt{s}}{2}n^\mu + \eta\frac{\sqrt{s}}{2}\bar{n}^\mu + k_\bot^\mu , 
\nonumber \\
k_2^\mu = \eta\frac{\sqrt{s}}{2}n^\mu + \bar{\eta}\frac{\sqrt{s}}{2}\bar{n}^\mu - k_\bot^\mu ,
\label{k12}
\end{eqnarray}
where, at large momentum transfer, $\eta$,  $\bar{\eta}$ and $k_\bot$ can be determined from
\begin{eqnarray}
\eta \simeq -\frac{t}{s}, \qquad \bar{\eta} = 1 - \eta \simeq -\frac{u}{s},  \qquad
k_\bot^2 \simeq \eta\bar{\eta} s,
\label{eta}
\end{eqnarray}
with the obvious restriction $0<\eta<1$. 
Note that the kinematic variable $\eta$ can be expressed in terms the $e^-$ c.m. angle $\theta$  as~:
\begin{eqnarray}
\eta \simeq \frac{1}{2}(1+\cos\theta)
\end{eqnarray}

The proton matrix element at leading twist level is described by twist-three nucleon DAs as:
\begin{equation}
~4\left\langle 0\left\vert \varepsilon^{ijk} u_{\alpha}W^{i}[\lambda_{1}n]u_{\beta}W^{j}[\lambda
_{2}n]d_{\sigma}W^{k}[\lambda_{3}n]\right\vert p_{1}\right\rangle
=\int Dx_{i}~e^{-ip_{1+}\left(  \sum x_{i}%
\lambda_{i}\right)  }{\Psi_{\alpha\beta\sigma}}(x_{i}), \label{DA:def}%
\end{equation}
with measure given by $\ ~Dx_{i}=dx_{1}dx_{2}dx_{3}\delta(1-x_{1}-x_{2}-x_{3})$, 
and where 
\begin{equation}
q_{\alpha}W[x]\equiv q_{\alpha}(x)\text{P}\exp\left\{  ig\int_{-\infty}%
^{0}dt~(n\cdot A)(x+tn)\right\}  .
\end{equation}
Following Ref.~\cite{Chernyak:1983ej},  
the function $\Psi_{\alpha\beta\sigma}(x_{i})$ can be expressed  as~:  
\begin{align}
~\ \Psi_{\alpha\beta\sigma}(x_{i}) &  =V(x_{i})~p_{1+}\left[  {\scriptstyle\frac{1}{2}%
}\Dbn~C\right]  _{\alpha\beta}\left[  \gamma_{5}N^{+}\right]  _{\sigma
}+A(x_{i})~p_{1+}\left[  {\scriptstyle\frac{1}{2}}\Dbn \gamma_{5}C\right]
_{\alpha\beta}\left[  N^{+}\right]  _{\sigma}\nonumber\\
&  ~\ \ \ \ \ \ \ \ \ \ \ \ \ \ \ \ \ \ \ \ \ \ \ \ \ \ \ \ +T(x_{i}%
)~p_{1+}\left[  {\scriptstyle\frac{1}{2}}\Dbn \gamma_{\bot}~C\right]
_{\alpha\beta}\left[  \gamma^{\bot}\gamma_{5}N^{+}\right]  _{\sigma},
\label{Psi:def}%
\end{align}
where $N^{+}\equiv \frac{\Dbn\Dn}{4}N$ represents the large  component of the nucleon spinor,  
$C$ is charge conjugation matrix: $C^{-1}\gamma_{\mu}C=-\gamma_{\mu}^{T}$, 
and the scalar functions $A,\ V,\ T$ stand for the nucleon DAs. 

The hard rescattering contribution to the time-like $2 \gamma$-exchange amplitudes 
$\delta \tilde{G}_{M}$, and $s/m_N^2 \tilde F_3$ can be obtained from the 
results for elastic ep-scattering \cite{Kivel:2009eg} using crossing relations for the hard perturbative subprocess. In our kinematics, expressed by Eqs.~(\ref{p12}, \ref{k12}, \ref{eta}), this leads to the following substitution:
\begin{equation}
Q^{2}\rightarrow -q^{2}-i\varepsilon, \ \zeta \rightarrow \eta,
\end{equation}
where $\zeta$ is the  kinematical parameter introduced in Ref.~\cite{Kivel:2009eg}. 
We then obtain for the time-like $2 \gamma$-exchange amplitudes~:
\begin{align}
\delta \tilde{G}_{M} (q^2, \eta)= -\frac{\alpha_{em}\alpha_{s}}{q^{4}}
\left( \frac{2\pi}{3}\right)  ^{2}  
\int \frac{d[y_{i}]}{y_1 y_2 \bar y_2}\frac{d[x_{i}]}{x_1 x_2 \bar x_2}  
\frac{4(2 \eta - 1)  x_2 y_2 \mathbf{\Phi}(y_{i}, x_{i}) }
{\left[ x_2 \bar \eta + y_2 \eta - x_2 y_2 \right]\left[ x_2 \eta + y_2 \bar \eta - x_2 y_2 \right]} ,
\label{eq:gm}
\\
\frac{s}{m_{N}^2} \tilde{F}_{3} (q^2, \eta) = \frac{\alpha_{em}\alpha_{s}}{q^{4}}\left(
\frac{2\pi}{3}\right)  ^{2}  
\int \frac{d[y_{i}]}{y_1 y_2 \bar y_2}\frac{d[x_{i}]}{x_1 x_2 \bar x_2}
\frac{2( x_2 \bar y_2 +\bar x_2  y_2)\mathbf{\Phi}(y_{i}, x_{i}) }
{\left[ x_2 \bar \eta + y_2 \eta - x_2 y_2 \right]\left[ x_2 \eta + y_2 \bar \eta - x_2 y_2 \right]} ,
\label{eq:f3}
\end{align}
where $\mathbf{\Phi} $ denotes the specific combination of the nucleon distribution amplitudes:
\begin{align}
\mathbf{\Phi}(y_{i}, x_{i})&=
 {Q_u}^2 \, \left[ (V^\prime + A^\prime)(V + A) + 4 T^\prime T \right](3,2,1)
    \nonumber \\  &
 + Q_u Q_d   \left[ (V^\prime + A^\prime)(V + A) + 4 T^\prime T \right](1,2,3)
+2 Q_u Q_d  \left[ V^\prime V + A^\prime A \right](1,3,2), 
\end{align}
and the numbers in the brackets define the order of the momentum fractions 
in the arguments of the DAs: $V^\prime V(3,2,1)\equiv V^\prime(y_{3},y_{2},y_{1})V(x_{3},x_{2},x_{1})$. We also introduced the  quark charges $Q_u = +2/3$, $Q_d = -1/3$, 
the fine structure coupling $\alpha_{em} = e^2 / (4 \pi)$, 
and the QCD coupling $\alpha_s$. 

In general, the time-like amplitudes are  complex functions.  At tree level, the  expressions 
of Eqs.~(\ref{eq:gm})  and (\ref{eq:f3}) do not contain an imaginary part explicitly. 
This can be simply understood~: the $s$-channel cut requires the on-shell photons (see e.g. diagram in Fig.\ref{Fig:diagrams}) but at large $q^{2}$  their propagators are highly virtual and hence the tree amplitudes are real.  Therefore we can obtain nontrivial imaginary  contributions only from the loop corrections. In particular, computing leading logarithms associated with the renormalization of DAs and QCD coupling $\alpha_{s}$ one obtains imaginary contributions generated by time-like logarithms:  $\ln[-q^{2}-i\varepsilon]=\ln[q^{2}]-i\pi$.  Such effects can be easily accounted for by using 
the well known  formula for the analytic continuation of $\alpha_{s}$ \cite{Radyushkin99}:
\begin{equation}
\alpha_{s}(-q^{2})=\frac{\alpha_{s}(q^{2})}{1-i\beta_{0}\alpha_{s}(q^{2})/4}+...,
\label{alf}
\end{equation}
where $\beta_{0}=11-2/3n_{f}$ is the first term of the $\beta$-function. Eq.~(\ref{alf}) includes resummed large corrections 
$\sim \beta_{0} \alpha_{s}$ which can be important at intermediate energies where $\alpha_{s}$ is not too small. Similarly, solving the renormalization group equation, one obtains an imaginary part originating from the evolution of DAs. However, the resulting imaginary contributions provide quite small numerical  effects  for the regions of $q^{2}$ which we are going to discuss below, see e.g. Ref.~\cite{Bakulev:2000uh}.    

As can be seen from Eqs.~(\ref{eq:gm}, \ref{eq:f3}) the leading behavior of the amplitudes $\delta \tilde G_M$ and $s /m_N^2 \tilde F_3$ goes as $1/q^4$, whereas $\delta \tilde F_2$ is suppressed in the large momentum transfer limit, since it behaves as $1/q^6$.  We may expect that at intermediate energies $\sim 5-10~$GeV$^{2}$ the effective scale defining the applicability of the perturbative expansion is already large enough in order to apply the present formalism. In what follows, we assume that the scale of the strong coupling in Eqs.~(\ref{eq:gm}, \ref{eq:f3}) is of order 
$\mu^{2}_{R}\simeq 0.6~q^{2}$.   

To evaluate the convolution integrals given in Eqs.~(\ref{eq:gm}, \ref{eq:f3}), we need to consider a model description for the twist-3 DAs. In Ref.~\cite{Braun:2000kw} the asymptotic behavior of the DAs and their first conformal moments can be found as
\begin{eqnarray}
V(x_{i})   &\simeq& 120x_{1}x_{2}x_{3} f_{N}\left[  1+r_{+}(1-3x_{3})\right], \nonumber \\
A(x_{i})   &\simeq& 120x_{1}x_{2}x_{3} f_{N}~r_{-}(x_{2}-x_{1}), \nonumber\\
T(x_{i})  &\simeq& 120x_{1}x_{2}x_{3} f_{N}\left[  1+\frac{1}{2}\left(
r_{-}-r_{+}\right)  (1-3x_{3})\right]  , \label{eq:DA}
\end{eqnarray}
where the DAs depend on the three parameters $f_N$, $r_+$ and $r_{-}$. For our calculations, we consider two phenomenological models for the DAs, which have been discussed in the literature~: 
COZ~\cite{Chernyak:1987nu} and BLW~\cite{Braun:2006hz}, as well as one description based 
on lattice QCD calculations \cite{Gockeler:2008xv}. The corresponding parameters are presented in Table \ref{table:DAmodels}. One notices that the parameters $r_+$ und $r_{-}$ in the BLW model and in the lattice calculations are nearly comparable, whereas the overall normalization $f_N$ is about 2/3 smaller for the lattice DA as compared with the description of the BLW model. In contrast to the BLW model and lattice calculations, the parameters $r_+$ and $r_{-}$ are about 3 times larger in the COZ description of the nucleon DAs. 
Below, we will provide calculations using the first two models, COZ and BLW. The results following from the lattice calculations can easily be approximated by scaling the BLW results.   All parameters from the Table~\ref{table:DAmodels} have been evolved  with leading logarithmic accuracy.  
\begin{center}
\begin{table}
\begin{tabular}{|c|c|c|c|}
\hline
& $f_N $ & $r_-$   & $r_+$       \\
& ($10^{-3}$ GeV$^2$) &  &  \\
\hline \; COZ \cite{Chernyak:1987nu} \; &
$5.0 \pm 0.5$ & $ 4.0 \pm 1.5 $ & $ 1.1 \pm 0.3 $   \\
\hline \; BLW \cite{Braun:2006hz} \; & 
$5.0 \pm 0.5$ & $1.37$ & $0.35 $ \\
\hline \; QCDSF \cite{Gockeler:2008xv}  \; & 
$3.23$ & $1.06$ & $0.33 $ \\
\hline
\end{tabular}
\caption{Parameters entering the proton DA (at $\mu$ = 1~GeV) 
for three parameterizations (COZ, BLW, and the lattice evaluation from QCDSF) 
used in this work.}
\label{table:DAmodels}
\end{table}
\end{center}

Using the parametrization of Eq.~(\ref{eq:DA}), the convolution integrals can be computed and yield~: 
\begin{align}
&\delta \tilde{G}_M(q^{2},\eta) = -\left(\frac{2\pi}{3}\right)^2(120f_N)^2 \frac{8}{9}\frac{\alpha_{em}\alpha_s}{q^4}
\nonumber  \\
&\ \ \ \ \ \ \ \ \ \ \ \  \times \ \Bigg\{ (\Phi_1 + \Phi_0)\Big[(\bar{\eta} - \eta)\eta\bar{\eta}\ln^2\left(\frac{\bar{\eta}}{\eta}\right) -4\eta\bar{\eta}\ln\left(\frac{\bar{\eta}}{\eta} \right) + (\eta-\bar{\eta})(1-\eta\bar{\eta}\pi^2)   \Big] 
\\
& + \Phi_2\Big[-3(\eta\bar{\eta})^2(\eta-\bar{\eta})\ln^2\left(\frac{\bar{\eta}}{\eta}\right) + \eta\bar{\eta}(1-12\eta\bar{\eta})\ln\left(\frac{\bar{\eta}}{\eta}\right)  + 3(\eta-\bar{\eta})\eta\bar{\eta}(1-\eta\bar{\eta}\pi^2) + \frac{1}{4}(\eta-\bar{\eta})  \Big]\Bigg\},  \nonumber 
\end{align}
\begin{align}
&\frac{s}{m^{2}_{N}} \tilde{F}_3(q^{2},\eta) = \left(\frac{2\pi}{3}\right)^2(120f_N)^2\frac{8}{9} \frac{\alpha_{em}\alpha_s}{q^4} 
\nonumber   \\
&\ \ \ \ \ \ \ \ \ \ \ \    \times\ \Bigg\{ -2(\Phi_1 + 2\Phi_0)\Big[(\eta-\bar{\eta})\ln\left(\frac{\bar{\eta}}{\eta}\right) -\eta\bar{\eta}\ln^2\left(\frac{\bar{\eta}}{\eta} \right) + 1 -\eta\bar{\eta}\pi^2 \Big]  \\
&  \ + \ \Phi_2\Big[-2\eta\bar{\eta}(1-6\eta\bar{\eta})\ln^2\left(\frac{\bar{\eta}}{\eta}\right) -12\zeta\bar{\eta}(2\eta-1)\ln\left(\frac{\bar{\eta}}{\eta}\right) 
 + 1 -12\eta\bar{\eta} - 2\eta\bar{\eta}(1-6\eta\bar{\eta})\pi^2 \Big] \Bigg\}, \nonumber
\end{align}
where the notation $\Phi_i$ denotes the following combinations of parameters:
\begin{eqnarray}
\qquad \Phi_0 &=& \frac{3}{4} + \frac{1}{2}r_{-} - \frac{1}{9}r_{-}^2 - \frac{3}{2}r_+ - r_+^2 + \frac{1}{3}r_+r_{-},  \nonumber \\
\qquad \Phi_1 &=& \frac{1}{2}r_{-} + \frac{1}{9}r_{-}^2 + \frac{3}{2}r_+ + \frac{5}{2}r_{+}^2 - \frac{5}{6}r_+r_{-},  \nonumber \\
\qquad \Phi_2 &=& \frac{7}{18}r_{-}^2 -\frac{11}{2}r_+^2 + \frac{4}{3}r_+r_{-}. \nonumber
\end{eqnarray}

\section{Results and Discussion}
\label{sec:results}

We calculate the $2 \gamma$-contribution to the cross section $\delta_{2\gamma}$, which is defined by
\begin{equation}
d\sigma_{\mathrm{c.m.}} = d\sigma_{\mathrm{c.m.},1\gamma}\left( 1 + \delta_{2\gamma} \right),
\end{equation}
as a function of the c.m. scattering angle $\theta$, where the cross section $d\sigma$ is given by Eq. (\ref{eq:crosssec}) and the cross section in Born approximation $d\sigma_{\mathrm{c.m.},1\gamma}$ by Eq. (\ref{cs:1photon}).

In order to estimate the relative effect of the $2 \gamma$-exchange in the unpolarized cross-section we need the input for the time-like FFs in Born approximation, $G_M$ and $G_E$.
We first start with a simple description of $G_M$, given by
\begin{equation}
|G_M(q^2)| = \frac{C}{q^4\log^2\left(\frac{q^2}{\Lambda^2} \right)},
\label{eq:qcdfit}
\end{equation}
with $\Lambda=0.3\mbox{ GeV}$ and with $C$ a free fit parameter. 
Furthermore, for the first FF parameterization, we use the assumption $|G_M| = |G_E|$ and neglect the imaginary part of the FFs in our calculation.

Since the value of $\delta\tilde F_2$ is unknown, we make an estimate of $\delta \tilde G_E$ using a simple model
\begin{equation}
\delta \tilde G_E \simeq \lambda\, \delta\tilde G_M , 
\label{eq:lambda}
\end{equation}
where $\lambda$ is a numerical parameter with $-1<\lambda<1$.

In our numerical calculations we fix the scale of the running coupling to be $\mu^2 = 0.6\ q^2$ due to the observation that the scale of the QCD strong coupling is usually lower than the value of $q^2$.
\begin{figure}
\includegraphics[width=8.0cm]{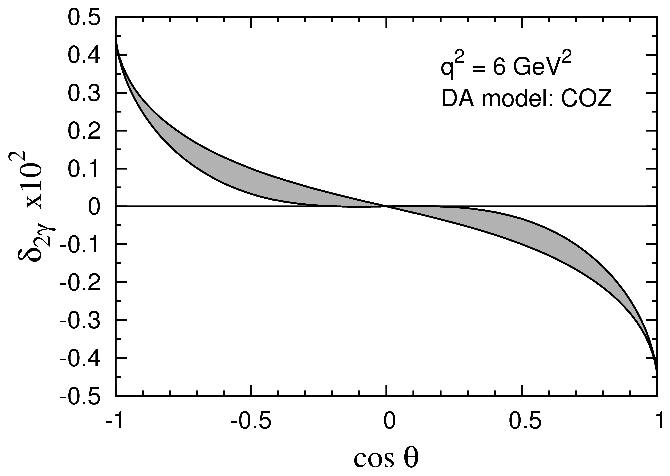}
\includegraphics[width=8.0cm]{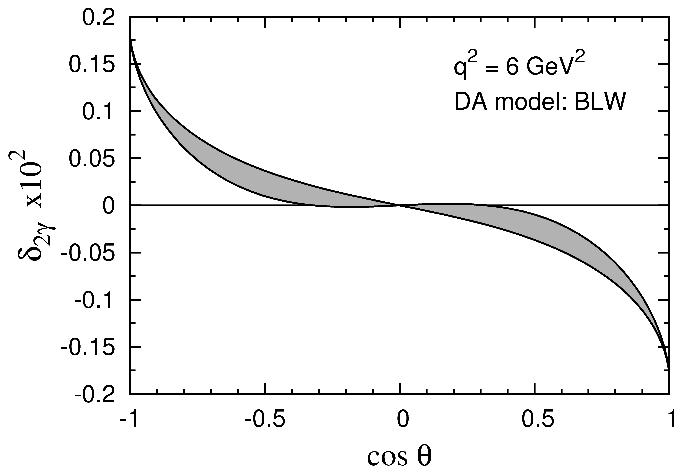}
\caption{Relative $2 \gamma$-contribution to the cross section at $q^2=6\ \mathrm{GeV}^2$ as a function of $\cos\theta$. The left (right) panels correspond to the calculations using the COZ (BLW) model for the proton DAs respectively. The bands describe the contribution for different values of $\delta \tilde G_E$ given by $-\delta \tilde G_M<\delta \tilde G_E <\delta \tilde G_M$.}
\label{fi:angledep}
\end{figure}

In Fig.~\ref{fi:angledep} we show the relative $2 \gamma$-contribution to the cross section for $s= 6\ \mathrm{ GeV}^2$ as a function of $\cos\theta$ using the two different models for the proton DAs. The shaded region corresponds to the variation of the parameter $\lambda$ in Eq.~(\ref{eq:lambda}). We observe that for both models the relative effect is smaller than 1\%. In both cases the 
angular dependence is similar, whereas the COZ model leads to a contribution which is twice as large as when using the BLW model.

\begin{figure}
 \includegraphics[width=8.0cm]{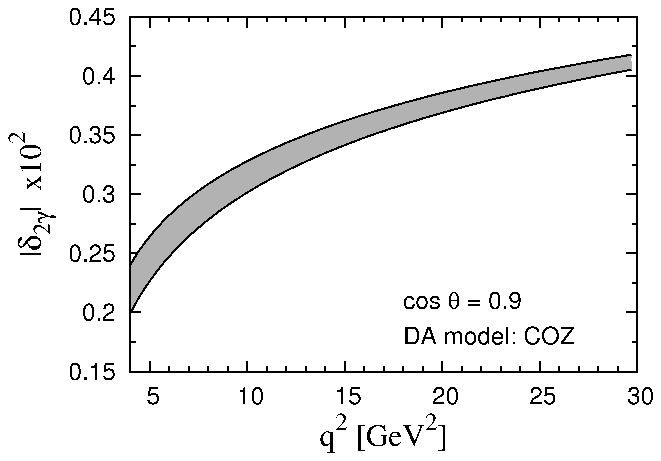}
\includegraphics[width=8.0cm]{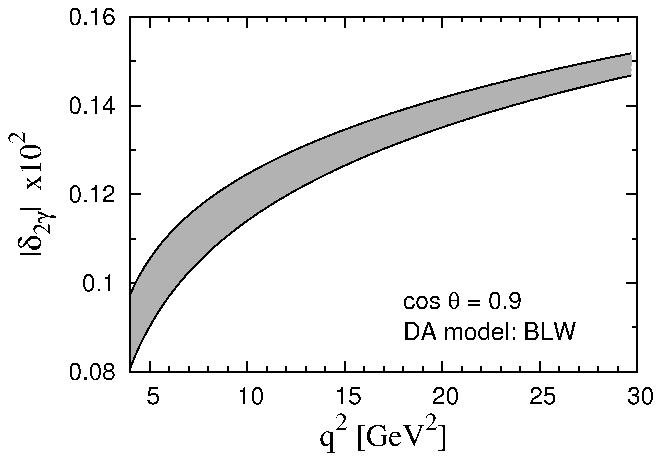}
\caption{The magnitude of the relative $2 \gamma$-exchange contribution as a function of $q^2$ for $ \cos\theta = 0.9$. The left (right) panels indicate the contribution calculated with the COZ (BLW) model 
respectively. The bands describe the contribution for different values of $\delta \tilde G_E$ 
given by $-\delta \tilde G_M<\delta \tilde G_E <\delta \tilde G_M$.}
\label{fi:qdep}
\end{figure}
The dependence of the $2 \gamma$-contribution on the momentum transfer $q^2$ is shown in 
Fig.~\ref{fi:qdep}. The absolute value of the correction is increasing with $q^2$, but this growth is logarithmic and can not change the effect quantitatively.

\begin{figure}
\centering \includegraphics[width=8.0cm]{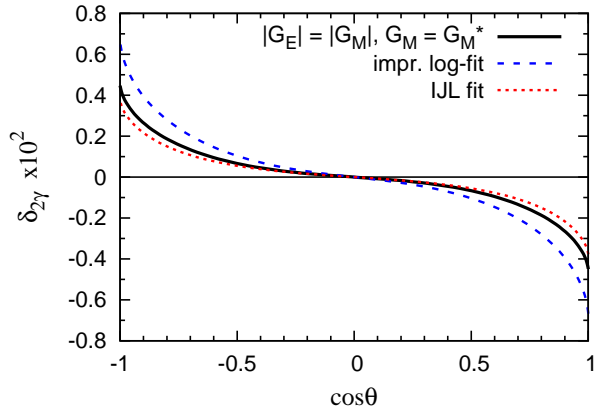}
\caption{Relative $2 \gamma$-exchange contribution for $q^2 = 6\ \mathrm{GeV}^2$ and $\delta\tilde G_E=0$ calculated with the COZ model using different parametrizations of the FFs. Solid black curve: using the assumptions $|G_E|=|G_M|$ and $G_M=G_M^*$ (purely real FF). Dashed blue curve: improved fit of Ref.~\cite{Brodsky:2003gs} involving logarithms for $F_2/F_1$. Dotted red curve: two-component fit of the FFs as represented in \cite{Iachello:2004aq}.}
\label{fi:modeldep}
\end{figure}
In addition we use two further models to parametrize the FFs entering the $1 \gamma$-amplitude and compare the results with the contribution we obtained using the simple fit given by Eq.~(\ref{eq:qcdfit}). Following \cite{Brodsky:2003gs}, we first consider an improved fit of the ratio $F_2/F_1$, which includes logarithmic corrections to the power law fall-off expected from QCD.
Furthermore we use a two-component fit of the FFs of Ref.~\cite{Iachello:2004aq}.
The results are demonstrated in Fig. \ref{fi:modeldep}, where $\delta_{2\gamma}$ has been calculated for $q^2=6\ \mathrm{GeV}^2$ with the COZ model for the nucleon DAs. All parametrizations lead to a similar behavior of the $2 \gamma$-contribution with respect to the scattering angle and to comparable quantitative results.

Furthermore we also calculated the single spin asymmetry $A_y$, which is given by Eq.~(\ref{eq:ssa}), for the two different parametrization of the FFs in Born approximation mentioned above, \cite{Brodsky:2003gs, Iachello:2004aq}. As discussed in Sec.~\ref{sec:calc}, we obtain a small imaginary part of the $2\gamma$ amplitudes $\delta \tilde G_M$ and $s/m_N^2 \tilde F_3$ in this approach, therefore the $2\gamma$ contribution to the single spin asymmtry mostly results from the interference of the real part of the $2\gamma$-amplitudes and the imaginary part of the FFs $G_E$ and $G_M$. We find that the relative contribution to $A_y$ is small as well, the effect is of order of about $1\%$.
\section{Conclusions}
\label{sec:conclusion}

In this work we provided a first calculation of the $2 \gamma$-exchange corrections in $p \bar{p} \rightarrow e^+e^{-}$ using a pQCD factorization approach. We obtain a small contribution to the cross section of $\lesssim \ 1\%$, in the momentum transfer range 5 - 30~GeV$^2$.  
The small $2 \gamma$-effect makes it challenging to observe such effects in unpolarized cross section measurements, e.g. by PANDA, see \cite{Sudol:2009vc}.
Since the value of the $2 \gamma$-contribution is sensitive to the choice of the DAs, a measurement of the process would allow to probe proton and antiproton DAs.

\section*{Acknowledgments}
The work of J.G. was supported by the Research Centre ``Elementarkraefte
und Mathematische Grundlagen" at the Johannes Gutenberg University
Mainz. The authors like to thank 
F. Maas and M. Sudol for helpful discussions. 


\end{document}